# Modulation and nonlinear evolution of multi-dimensional Langmuir wave envelopes in a relativistic plasma

**M. Shahmansouri**
*Department of Physics, Faculty of Science, Arak University, Arak 38156- 8 8349, Iran.*

**A. P. Misra**
*Department of Mathematics, Siksha Bhavana, Visva-Bharati University, Santiniketan-731 235, India.*

The modulational instability (MI) and the evolution of weakly nonlinear two-dimensional (2D) Langmuir wave (LW) packets are studied in an unmagnetized collisionless plasma with weakly relativistic electron flow. By using a 2D self-consistent relativistic fluid model and employing the standard multiple-scale technique, a coupled set of Davey-Stewartson (DS)-like equations is derived which governs the slow modulation and the evolution of LW packets in relativistic plasmas. It is found that the relativistic effects favor the instability of LW envelopes in the $k-\theta$ plane, where $k$ is the wave number and $\theta$ $(0 \leq \theta \leq \pi)$ the angle of modulation. It is also found that as the electron thermal velocity or $\theta$ increases, the growth rate of MI increases with cutoffs at higher wave numbers of modulation. Furthermore, in the nonlinear evolution of the DS-like equations, it is seen that with an effect of the relativistic flow, a Gaussian wave beam collapses in a finite time, and the collapse can be arrested when the effect of the thermal pressure or the relativistic flow is slightly relaxed. The present results may be useful to the MI and the formation of localized LW envelopes in cosmic plasmas with a relativistic flow of electrons.

## I. INTRODUCTION

Laboratory [1] and space [2,3] observations indicate that there have been plenty of evidences for the existence of Langmuir waves (LWs) as the high-frequency oscillations in plasmas. These waves are associated with broadband electrostatic noise which is a common high-frequency background activity, regularly observed by satellite missions in the plasma sheet boundary layers. Also, the FAST satellite observations support the occurrence of LWs associated with auroral density cavities [4] and in the mid-altitude auroral zone [5]. Over the last many years, the propagation of LWs received a great deal of attention [6-20] for research due to their potential application in different plasmas environments.

In plasmas, the localization of nonlinear waves may occur in the form of an envelope pulse confining (modulating) carrier waves as governed by nonlinear Schrödinger (NLS)-like equations or solitons which are described by the Korteweg- de Vries (KdV)-like equations. These envelope solitons in plasmas are generically subject to their amplitude modulation due to self-interactions of plasma carrier waves [11, 13, 14, 17, 21-31] (i.e., a slow variation of the wave envelope due to the nonlinearities). The formation of such envelope solitons is a result of the balance between the nonlinearity (self-focusing) and the wave dispersion. Indeed, the dynamical occurrence of this type of balance can predicts the existence of solitons even under the strong perturbation. It must be added that the multidimensional NLS equation or Davey-Stewartson (DS)-like equations including the nonlocal nonlinearities may no longer be integrable, which may lead to the formation of the wave collapse or blowup of the wave amplitude due to self-focusing. Several attempts have been made to extend the theories and to investigate the multidimensional evolution of the wave envelopes in plasmas [23,26,32-39].

In this work, we consider a self-consistent 2D fluid model that describes the dynamics of weakly relativistic electrons in unmagnetized plasmas. We show that the 2D evolution of LW packets can be governed by a DS-like equation in plasmas with weakly relativistic flow of electrons. The effects of the relativistic thermal pressure and the obliqueness of modulation on the MI and the associated growth rate are studied. We also show that a Gaussian wave beam can self-focus and blow-up in a finite time. This singularity, can, however be removed when the effects of the thermal pressure are relatively low.

## II. THEORETICAL MODEL

We consider the propagation of weakly nonlinear LWs in a 2D unmagnetized plasma with weakly relativistic flow of electrons. The ion species are considered as a uniform stationary background in the electron-acoustic (EA) time scale. The weakly relativistic fluid equations are [40]

$$\frac{\partial n_e}{\partial t} + \nabla \cdot (n_e \mathbf{v}_e) = 0, \tag{1}$$

$$m_e \left(\frac{\partial}{\partial t} + \mathbf{v}_e \cdot \nabla\right)(\gamma(\mathbf{v}_e)) = -e\mathbf{E} - \frac{\gamma^2}{n_e}\left(\nabla + \frac{\mathbf{v}_e}{c^2}\frac{\partial}{\partial t}\right)P_e, \tag{2}$$

$$\nabla^2 \phi = n_e - n_0, \tag{3}$$

$$P_e = \left(\frac{n_e}{\gamma}\right)^3 \frac{T_e}{n_0^2}, \tag{4}$$

where $m_e$, $n_e$, $n_0$, $\mathbf{v}_e \equiv (v_x, v_y)$, $\varphi, \gamma$, and $P_e$ are, respectively, the electron mass, the number density, the equilibrium electron (or ion) number density, the electron fluid velocity, the electrostatic potential, the relativistic factor given by $\gamma = 1/\sqrt{1 - v_e^2/c^2}$, and the electron pressure (with $T_e$ denoting the electron temperature in energy units) including the relativistic electron flow [40].

To investigate the amplitude modulation of LWs, we adopt the standard multiple-scale technique [41] and obtain the Davey-Stewartson (DS) like- equation. Thus, the independent variables are stretched as

$$\xi = \varepsilon(x - v_{gx}t), \ \eta = \varepsilon(y - v_{gy}t), \ \tau = \varepsilon^2 t, \tag{5}$$

where $\varepsilon$ is a small dimensionless parameter measuring the weakness of nonlinearity and $v_{gx}$ ($v_{gy}$) is the $x$ ($y$)-component of the group velocity, to be determined later by the compatibility condition. The dependent variables (in which the perturbed parts depend on the fast scales via the phase $kx - \omega t$ and the slow scales enter only the $l$ th harmonic amplitude) are expanded as

$$F = F^{(0)} + \sum_{m=1}^{\infty} \varepsilon^m F^{(m)}, \tag{6a}$$

$$F^{(m)} = \sum_{l=0}^{m} \varepsilon^m F_l^{(m)}(\xi,\eta,\tau) \exp[il(kx - \omega t)] + c.c., \tag{6b}$$

where $F^{(0)} = n_0$ for $n_e$, while for other variables $F^{(0)} = 0$, c.c. denotes the complex conjugate; $\omega$ is the frequency and $k$ is the wave number of LWs. All the variables should satisfy the reality condition $F_l^{(m)} = F_{-l}^{(m)*}$ with asterisk denoting the complex conjugate.

In what follows, we substitute the stretched coordinates and the expansion given by Eqs. (5) and (6) into Eqs. (1)-(4), and collect the terms in different powers of $\varepsilon$.

In the lowest power of $\varepsilon$, i.e., for $m = 1$, $l = 1$, the first order harmonics yield the following dispersion relation for Langmuir waves [42]

$$\omega^2 = \omega_{pe}^2 + \frac{3}{2}k^2 V_{th}^2, \tag{7}$$

where $V_{th} = \sqrt{2T_e/m_e}$ is the electron thermal speed, and $\omega_{pe} = \sqrt{4\pi n_0 e^2/m_e}$ is the electron plasma oscillation frequency.

For the second order reduced equations with ($m = 2$) and ($l = 1$) we obtain the following compatibility condition for the group velocity components:

$$v_{gx} \equiv \partial\omega/\partial k_x = \frac{3k_x}{2\omega}V_{th}^2, \tag{8a}$$

$$v_{gy} \equiv \partial\omega/\partial k_y = \frac{3k_y}{2\omega}V_{th}^2. \tag{8b}$$

The second order ($m = 2$) with the second ($l = 2$) harmonics lead to the following expressions,

$$n_2^{(2)} = A_1\left(\varphi_1^{(1)}\right)^2, \tag{9a}$$

$$\varphi_2^{(2)} = A_2\left(\varphi_1^{(1)}\right)^2, \tag{9b}$$

$$k \cdot v_2^{(2)} = A_3\left(\varphi_1^{(1)}\right)^2, \tag{9c}$$

where the coefficients $A_i (i = 1,2,3)$ are given in Appendix A.

It is known that the zeroth harmonic components appear due to the nonlinear self-interaction of the carrier waves. Thus, the first order, zeroth harmonic components are obtained from the terms corresponding to ($m = 2$) and ($l = 0$) as

$$n_0^{(1)} = 0, \tag{10a}$$

$$\nabla_1 \mathbf{v}_0^{(1)} = -\frac{e}{m_\circ}\nabla_0 \varphi_0^{(1)}, \tag{10b}$$

$$\nabla_0 \cdot \mathbf{v}_0^{(1)} = 0, \tag{10c}$$

where $\nabla_0 = \hat{\mathbf{i}}\partial/\partial\xi + \hat{\mathbf{j}}\partial/\partial\eta$ and $\nabla_1 = v_{gx}\partial/\partial\xi + v_{gy}\partial/\partial\eta$. Here, we suppose that all the first order zeroth harmonic components are zero [43]. Also, the following equations are obtained for $m = 3$ and $l = 0$ for the second order, zeroth harmonic components of the velocity as

$$\chi_1 + v_{y0}^{(2)} = \frac{k^2}{4\pi m_\circ n_\circ \omega}\left[k_x \chi_6 + k_y \left|\phi_1^{(1)}\right|^2\right], \tag{11a}$$

$$v_{x0}^{(2)} + \chi_2 = \frac{k^2}{4\pi m_\circ n_\circ \omega}\left[k_x \left|\phi_1^{(1)}\right|^2 + k_y \chi_6\right], \tag{11b}$$

$$v_{gx} v_{x0}^{(2)} + v_{gy} \chi_3 = -\frac{e}{m_\circ}\phi_0^{(2)} + \alpha \left|\phi_1^{(1)}\right|^2 + \kappa\chi_6, \tag{11c}$$

$$v_{gx} \chi_4 + v_{gy} v_{y0}^{(2)} = -\frac{e}{m_\circ}\phi_0^{(2)} + \beta \left|\phi_1^{(1)}\right|^2 + \kappa\chi_5, \tag{11d}$$

where

$$\chi_1 = \int \frac{\partial v_{x0}^{(2)}}{\partial \xi} d\eta, \quad \chi_2 = \int \frac{\partial v_{y0}^{(2)}}{\partial \eta} d\xi, \quad \chi_3 = \int \frac{\partial v_{x0}^{(2)}}{\partial \eta} d\xi, \quad \chi_4 = \int \frac{\partial v_{y0}^{(2)}}{\partial \xi} d\eta, \quad \chi_5 = \int \frac{\partial}{\partial \xi}\left|\phi_1^{(1)}\right|^2 d\eta,$$

$\chi_6 = \int \frac{\partial}{\partial \eta}\left|\phi_1^{(1)}\right|^2 d\xi$, $\nabla_2 = k_x \partial/\partial\xi + k_y \partial/\partial\eta$, and the coefficients $\alpha$, $\beta$ and $\kappa$ are given in Appendix A.

Furthermore, the third order ($m = 3$), first harmonic ($l = 1$) components are obtained as

$$\frac{\partial n_1^{(1)}}{\partial \tau} - \nabla_1 n_1^{(2)} + \nabla_0 \cdot \mathbf{v}_1^{(2)} - i\omega n_1^{(3)} + i\mathbf{k}\cdot[\mathbf{v}_1^{(3)} + \mathbf{v}_2^{(2)} n_1^{(1)*} + \mathbf{v}_1^{(1)*} n_2^{(2)} + \mathbf{v}_1^{(1)} n_0^{(2)} + \mathbf{v}_0^{(2)} n_1^{(1)}] = 0, \tag{12a}$$

$$\frac{\partial \mathbf{v}_1^{(1)}}{\partial \tau} - \nabla_1 \mathbf{v}_1^{(2)} - i\omega \mathbf{v}_1^{(3)} + 2i(\mathbf{k}\cdot\mathbf{v}_1^{(1)*})\mathbf{v}_2^{(2)} + i(\mathbf{k}\cdot\mathbf{v}_0^{(2)*})\mathbf{v}_1^{(1)}$$
$$-i\omega\left[3(v_{x1}^{(1)})^2 v_{x1}^{(1)*}\hat{\mathbf{i}} + (v_{x1}^{(1)})^2 v_{y1}^{(1)*}\hat{\mathbf{j}} + 2v_{x1}^{(1)} v_{x1}^{(1)*} v_{y1}^{(1)}\hat{\mathbf{j}} + (v_{y1}^{(1)})^2 v_{x1}^{(1)*}\hat{\mathbf{i}} + 2v_{x1}^{(1)} v_{y1}^{(1)} v_{y1}^{(1)*}\hat{\mathbf{i}} + 3(v_{y1}^{(1)})^2 v_{y1}^{(1)*}\hat{\mathbf{j}}\right]$$
$$= \frac{e}{m_\circ}(i\mathbf{k}\phi_1^{(3)} + \nabla_0 \phi_1^{(2)}) \tag{12b}$$
$$-\frac{3T_e}{n_0 m_\circ}\left[i\mathbf{k} n_1^{(3)} - \frac{i\mathbf{k}}{2c^2}(2n_1^{(1)}\left|v_{x1}^{(1)}\right|^2 + n_1^{(1)*}(v_{x1}^{(1)})^2 + 2n_1^{(1)}\left|v_{y1}^{(1)}\right|^2 + n_1^{(1)*}(v_{y1}^{(1)})^2) + \nabla_0 n_1^{(2)}\right.$$
$$\left.+\frac{i\omega}{c^2}(n_1^{(1)*}\mathbf{v}_2^{(2)} - 2n_2^{(2)}\mathbf{v}_1^{(1)*} - (n_1^{(1)})^2 \mathbf{v}_1^{(1)*}) + i\mathbf{k}(n_2^{(2)} n_1^{(1)*} + n_0^{(2)} n_1^{(1)})\right]$$

$$-k^2 \phi_1^{(3)} + 2i\nabla_2 \phi_1^{(2)} + \nabla_0^2 \phi_1^{(1)} = 4\pi e n_1^{(3)}, \tag{12c}$$

Finally, we obtain the following coupled set of partial differential equations from Eqs. (11) and (12) for the propagation of modulated LWs:

$$i\frac{\partial \varphi}{\partial \tau} + P_1 \frac{\partial^2 \varphi}{\partial \xi^2} + P_2 \frac{\partial^2 \varphi}{\partial \eta^2} + P_3 \frac{\partial^2 \varphi}{\partial \xi \partial \eta} + Q_1 \varphi |\varphi|^2 + [Q_2 \chi_4 + Q_3 \chi + Q_4 \chi_5] \varphi = 0, \tag{13}$$

$$\frac{\partial \chi_4}{\partial \xi} + \frac{\partial \chi}{\partial \eta} = R_1 \frac{\partial |\varphi|^2}{\partial \xi} + R_2 \frac{\partial |\varphi|^2}{\partial \eta} + \frac{1}{k} R_3 \nabla_2 |\varphi|^2, \tag{14}$$

where $\varphi = \varphi_1^{(1)}$, $\chi = v_{y0}^{(2)}$ and $\psi = \chi_4$. The coefficients $P_i$ ($i = 1,2,3$), $Q_j$ ($j = 1,2,3,4$) and $R_k$ ($k = 1,2,3,4,5,6$) are given in Appendix A.

Thus, we have obtained a new set of nonlocal nonlinear equations, which describe the slow modulation of the LW packets in 2D plasmas with weakly relativistic electron flow. The coefficients $P_1$ and $P_2$ appear due to the LW group dispersion and the 2D motion, whereas $P_3$ is for the arbitrary orientations of the carrier waves as well as the modulation of the LW packets. The cubic nonlinear coefficient $Q_1$ is due to the carrier wave self-interaction originating from the zeroth harmonic modes (or slow modes), i.e., the ponderomotive force, and $Q_4$ is a result of the combined effects of 2D motion and self-interaction. The nonlinear nonlocal coefficients $Q_2$ and $Q_3$ appear due to the coupling between the dynamical field associated with the first harmonic (with a "cascaded" effect from the second harmonic) and a static field generated due to the mean motion (zeroth harmonic) in the relativistic plasma. The appearance of the coefficients $R_j, j = 1,2,3$ can also be explained similarly.

Looking for the modulation of LW packets parallel to the carrier wave vector, one can obtain the DS-like equations [44] by setting $k_y \to 0$ into Eqs (13) and (14) as

$$i\frac{\partial \varphi}{\partial \tau} + P_{10} \frac{\partial^2 \varphi}{\partial \xi^2} + P_{20} \frac{\partial^2 \varphi}{\partial \eta^2} + [Q_{10} |\varphi|^2 + Q_{20} \psi] \varphi = 0, \tag{15}$$

$$\frac{\partial^2 \psi}{\partial \xi^2} + \frac{\partial^2 \psi}{\partial \eta^2} = R_0 \frac{\partial^2}{\partial \xi^2} |\varphi|^2, \tag{16}$$

where $P_{j0} \equiv (P_j)|_{k_y \to 0}$, $Q_{j0} \equiv (Q_j)|_{k_y \to 0}$, $R_{j0} \equiv (R_j)|_{k_y \to 0}$ and $R_0 = R_{10} + R_{30}$.

Furthermore, in the case of $k_y \to 0$ and by disregarding the $\eta$-dependency of the physical variables the set of Eqs. (13) and (14) can be reduced to a well-known NLS equation as

$$i\frac{\partial \varphi}{\partial \tau} + P_1 \frac{\partial^2 \varphi}{\partial \xi^2} + [Q_{10} + Q_{20} R_0] |\varphi|^2 \varphi = 0, \tag{17}$$

### III. MODULATIONAL INSTABILITY ANALYSIS

We consider the modulation of a plane wave solution of Eqs. (15) and (16) for $\phi$ and $\psi$. We choose $\psi = \psi_0$ and a plan wave solution for $\varphi$ of the form $\phi = \phi_0 \exp[i\Delta(\tau)]$, where $\Delta = 2Q_{10}\varphi_0^2 + Q_{20}\psi_0$, and $\psi_0$, $\varphi_0$ are constants. Note that the stability criterion does not depend on the choice of $\psi = \psi_0$. Next, to study the stability of this solution, we modulate the wave amplitude against a plane wave perturbation as $\phi = (\phi_0 + \delta\phi)\exp[i\Theta(\xi,\eta,\tau)]$ and $\psi = \psi_0 + \delta\psi$, where $(\delta\phi, \delta\psi, \delta\Theta) \equiv (A, B, C)\exp(i\mathbf{K}\cdot\mathbf{R} - i\Omega\tau)$, $\mathbf{R} \equiv (\xi,\eta)$, and also ($\mathbf{K} \equiv (K_\xi, K_\eta)$) and ($\Omega$), respectively, denote the wave number and the wave frequency of modulation. Substituting the above perturbed solutions into Eqs. (15) and (16) we obtain the following nonlinear dispersion relation for the modulated LW packets

$$\Omega^2 = (P_{10}K_\xi^2 + P_{20}K_\eta^2)\left[P_{10}K_\xi^2 + P_{20}K_\eta^2 - 2Q_{10}\phi_0^2 - 2Q_{20}R_0\frac{K_\xi^2}{K_\xi^2 + K_\eta^2}\phi_0^2\right]. \quad (18)$$

This shows that the modulation of the amplitude of the electric potential $\phi$ typically depends on the coefficients $P_{1,20}$, $Q_{1,20}$ and $R_0$. Equation (18) can be rewritten as

$$\Omega^2 = (P_{10}K_\xi^2 + P_{20}K_\eta^2)^2\left[1 - \frac{K_c^2}{K^2}\right], \quad (19)$$

where $K_c$ is the critical wave number, given by

$$K_c^2 = \frac{2Q_{10}(1+\tan^2\theta) + 2Q_{20}R_0}{P_{10} + P_{20}\tan^2\theta}\phi_0^2, \quad (20)$$

with $\theta = \tan^{-1}(K_\eta/K_\xi)$. Equation (19) shows that the MI sets in for a wave number $K < K_c$ (i.e., for all wavelengths above the threshold $\lambda_c = 2\pi/K_c$). It turns out that the right-hand side of Eq. (20) must be positive, which leads to the following constraint

$$\Lambda = [Q_{10}(1+\tan^2\theta) + Q_{20}R_0](P_{10} + P_{20}\tan^2\theta) > 0. \quad (21)$$

On the other hand, for $K > K_c$, the LW packet is said to be stable under the modulation. The growth rate of instability (letting $\Gamma = |\text{Im}(\Omega(K))|$) can be obtained from Eq. (19) as

$$\Gamma = \frac{K^2(P_{10} + P_{20}\tan^2\theta)}{1 + \tan^2\theta}\sqrt{\frac{K_c^2}{K^2} - 1} \quad (22)$$

Thus, the maximum value of the growth rate can be reached at $K = K_c/\sqrt{2}$, i.e., $\Gamma_{max} = [Q_{10} + Q_{20}R_0\cos^2\theta]\phi_0^2$.

We numerically investigate the stable/unstable regions for the plane-wave solution relying on the above condition (21), which, in turn, typically depends on the electron thermal speed $V_{th}$ and the modulational obliqueness $\theta$. Figure 1 displays the contour plots of $\Lambda = 0$ in the $(k,\theta)$-plane for different values of $V_{th}$ to show the stable $(\Lambda < 0)$ and unstable $(\Lambda > 0)$ regions. The pitch angle $\theta$ is allowed to vary between $0$ and $\pi$, so that all the subplots are symmetric upon reflection with respect to either $\theta = 0$ or $\theta = \pi/2$ lines. In the numerical investigation, we have normalized the physical quantities as follows: $k \to k/\lambda_D$, $\omega \to \omega/\omega_{pe}$, $P_{1,2} \to P_{1,2}/(V_{th}^2/\omega_{pe})$, $Q_1 \rightarrow Q_1/(e^2\omega_{pe}/m_e^2 V_{th}^4)$, $Q_2 \rightarrow Q_2 V_{th}/\omega_{pe}$, $R_0 \rightarrow R_0 V_{th}^3 m_e^2/e^2$ where $\lambda_D = V_{th}/\omega_{pe}$ is the Debye length. In Fig. 1, the subplots (a), (b) and (c) are corresponding to $V_{th}/c = 0.2, 0.1, 0.05$ respectively. The blank or white and the shaded or colored areas indicate the regions in the *(k,θ)* plane where $\Lambda < 0$ and $\Lambda > 0$. It is seen that regardless of the values of $\theta$, the wave is stable in the very long-wavelength $(0 \leq k \leq 0.01)$ and short-wavelength $(k > 1)$ regions. For some other regions of $k$, the wave can be stable or unstable depending on the values of $\theta$ and the electron thermal velocity $V_{th}$. Furthermore, as the electron thermal velocity decreases or the relativistic effect becomes weaker, part of the unstable region shifts to a stable one. This means that the relativistic electron flow and the obliqueness of modulation favor the MI of LW packets. These are in agreement with the previous results where the acoustic waves are found to be unstable against the parallel modulation if one takes into account finite temperature effects [45-47] or considers an oblique modulation of the wave amplitude [48,49].

The growth rate of MI is plotted in Fig. 2 for different values of $V_{th}$ and the obliqueness parameter θ. It is found as these parameters increase, the value of $\Gamma$ increases having cutoffs at higher values of the wave number of modulation. Thus, the instability growth rate can be suppressed in nonrelativistic or weakly relativistic plasmas with 2D propagation of LWs.

Next, we examine the long-time evolution of the LW packets in relativistic plasmas. To this end, we perform a numerical simulation of Eqs. (15) and (16) in which the derivatives are discretized using a finite difference scheme, and considered are the domain size $25 \times 25$ with $150 \times 150$ grid points and time step $\tau = 10^{-3}$ for (2 + 1)-dimensional evolution. We choose the initial condition as a symmetric Gaussian wave beam of the form $\phi = \sqrt{2I/\pi ab}\exp(-\xi^2/a^2 - \eta^2/b^2)$, where $I$ is the wave action. The results are shown in Fig. 3 at different time intervals. It is seen that the wave beam self-focuses and blows up within a short time when the electron thermal velocity or the relativistic effect is relatively strong, namely $V_{th} = 0.2c$ and with $k = 0.1$. It turns out that the validity of Eqs. (15) and (16) breaks down near this singularity. We find that when the relativistic effect is bit relaxed or the electron thermal velocity is lowered, the collapse is arrested with a stable oscillation (See Fig. 4) for a long time, e.g., . $\tau = 100$.

## IV. CONCLUSONS

We have investigated the 2D propagation of weakly nonlinear LWs in an unmagnetized plasma with a flow of weakly relativistic electrons. Starting from a self-consistent relativistic fluid model and using the standard multiple scale technique, we have derived a DS-like equations [44] which govern the evolution of LW envelopes in relativistic plasmas. The oblique MI of LWs is then studied against a plane wave perturbation. We have shown that the electron thermal speed in relativistic flow have the significant role on the existence domain of instability in the *(k,θ)*-plane. It not only favors the MI, but also increases the corresponding growth rates with cutoffs at higher values of the wave number of modulation. Similar roles of the obliqueness parameter are also found. A numerical simulation of the DS equations reveals that a Gaussian wave beam can self-focus after interaction and blows up in a finite interval of time when the electron thermal velocity is relatively high and so is the MI growth. This singularity can, however, be removed with comparatively lower thermal velocity of electrons.

To conclude, our results may be applicable to the modulation and the evolution of Langmuir wave envelopes in plasmas with relativistic flow of electrons such as those in laboratory and cosmic plasma environments where $n_0 \sim 10^{10} - 10^{14}/cm^3$, $T_e \sim 10^6 - 10^8 K$ and $V_{th} \sim 0.01c - 0.1c$.

**Acknowledgments**

A. P. Misra is supported by UGC-SAP (DRS, Phase-III) with Sanction order No. F.510/3/DRS-III/2015 (SAPI) and UGC-MRP with F. No. 43-539/2014 (SR) and FD Diary No. 3668.

**References**

[1] D. Henry and J. P. Trguier, J. Plasma Phys. **8**, 311 (1972).

[2] W. C. Feldman, R. C. Anderson, S. J. Bame, S. P. Gary, J. T. Gosling, D. J. McComas, M. F. Thomsen, G. Paschmann, M. M. Hoppe, J. Geophys. Res. **88**, 96 (1983).
[3] S. D. Bale, P. J. Kellogg, D. E. Larson, R. P. Lin, K. Geotz, P. Lepping, Geophys. Res. Lett. **25**, 2929 (1998)
[4] R. Pottelette, R. E. Ergun, R. A. Treumanna, M. Berthomier, C. W. Carlson, J. P. McFadden, and I. Roth, Geophys. Res. Lett. **26**, 2629 (1999).
[5] R. E. Ergun, C. W. Carlson, J. P. McFadden, E. S. Moser, G. T. Delory, W. Peria, C. C. Chaston, M. Ternerin, I. Roth, L. Muschiet, R. Elphic, R. Strangeway, R. Pfaff, C. A. Cattell, D. Klumpar, E. Shelley, W. Peterson, E. Moebius, and L. Kistler, Geophys. Res. Lett. **25**, 2041 (1998).
[6] K. Watanabe and T. Taniuti, J. Phys. Soc. Jpn. **43**, 1819 (1977).
[7] M. Y. Yu and P. K. Shukla, Plasma Phys. **29**, 409 (1983).
[8] N. Dubouloz, R. Pottelette, M. Malingre, and R. Treumann, Geophys. Res. Lett. **18**, 155 (1991).
[9] S. V. Singh and G. S. Lakhina, Planet. Space Sci. **49**, 107 (2001).
[10] R. L. Mace and M. A. Hellberg, Phys. Plasmas **8**, 2649 (2001).
[11] P. K. Shukla, M. A. Hellberg, and L. Stenflo, J. Atmosph. Sol. Terrest. Phys. **65**, 355 (2003).
[12] S. G. Tagare, S. V. Singh, R. V. Reddy, and G. S. Lakhina, Nonlinear Processes Geophys. **11**, 215 (2004).
[13] I. Kourakis and P. K. Shukla, Phys. Rev. E **69**, 036411 (2004).
[14] C. Bhowmik, A. P. Misra, and P. K. Shukla, Phys. Plasmas **14**, 122107 (2007).
[15] T. S. Gill, H. Kaur, S. Bansal, N. S. Saini, and P. Bala, Eur. Phys. J. D **41**, 151 (2007).
[16] G. S. Lakhina, A. P. Kakad, S. V. Singh, and F. Verheest, Phys. Plasmas **15**, 062903 (2008).
[17] A. P. Misra and P. K. Shukla, Phys. Plasmas **15**, 122107 (2008).
[18] J. Borhanian and M. Shahmansouri, Astrophys. Space Sci. **342**, 401 (2012).
[19] P. Carbonaro, Chaos, Solitons & Fractals **45** 959 (2012).
[20] M. Shahmansouri and H. Alinejad, Astrophys. Space Sci. **347,** 305 (2013).
[21] A. P. Misra and P. K. Shukla, Phys. Plasmas **14**, 082312 (2007); Phys. Plasmas **15**, 052105 (2008).
[22] A. P. Misra, C. Bhowmik, and P. K. Shukla, Phys. Plasmas **16**, 072116 (2009).
[23] T. S. Gill, C. Bedi , and A. S. Bains, Phys. Plasmas **16**, 032111 (2009).
[24] A. P. Misra and A. R. Chowdhury, Phys. Plasmas **14**, 012309 (2007).
[25] A. S. Bains, M. Tribeche, and T. S. Gill, Phys. Lett. A **375**, 2059 (2011).
[26] H. Alinejad, M. Mahdavi, and M. Shahmansouri, Astrophys. Space Sci. **352**, 571 (2014).
[27] S. Chandra and B. Ghosh, Astrophys. Space Sci. **342**, 417 (2012).
[28] D-N. Gao, C-L. Wang, X. Yang, W-S. Duan, and L. Yang, Phys. Plasmas **19**, 122112 (2012).
[29] N. S. Saini and R. Kohli, Astrophys. Space Sci. **349**, 293 (2014).
[30] A. Merriche and M. Tribeche, Physica A **421**, 463 (2015).
[31] M. McKerr, F. Hass, and I. Kourakis, Phys. Plasmas **23**, 052120 (2016).
[32] M. R. Amin, G. E. Morfill, P. K. Shukla, Phys Rev E **58**, 6517 (1998).
[33] K. Nishinari, K. Abe, and Satsuma, J. Theoret Math Phys **99**, 745 (1994).
[34] S. S. Ghosh, A. Sen, and G. S. Lakhina, Nonlinear Processes Geophys **2002,** 463 (2002).

[35] W. S. Duan, Phys Plasmas **19**, 3022 (2003).
[36] Ju-Kui Xue, Phys. Plasmas **12**, 062313 (2005).
[37] Y. Y. Wang and J. F. Zahng, Phys lett A **372**, 6509 (2008).
[38] A. P. Misra and P. K. Shukla, Phys. Plasmas **18**, 042308 (2011).
[39] A. P. Misra, M. Marklund, G. Brodin, and P. K. Shukla1 Phys. Plasmas **18**, 042102 (2011).
[40] T. Katsouleas and W. B. Mori, Phys. Rev. Lett. **61**, 90 (1988); J. B. Rosenzweig, Phys. Rev. A **38**, 3634 (1988).
[41] T. Tanuiti and N. Yajima, J. Math. Phys. **10**, 1369 (1969).
[42] F. Chen, Introduction to Plas,ma Physics (New York: Springer. 1974)
[43] T. Taniuti, Suppl. Prog. Phys. **55**, 1 (1974).
[44] A. Davey and K. Stewartson, Proc. R. Soc. London, Ser. A **338**, 101 (1974).
[45] V. Chan and S. Seshadri, Phys. Fluids **18**, 1294 (1975).
[46] I. Durrani, G. Murtaza, U. H. Rahman, and I. A. Azhar, Phys. Fluids **22**, 791 (1979).
[47] J-K. Xue, W-S. Duan, and L. He, Chin. Phys. **11**, 1184 (2002).
[48] R. Chhabra and S. Sharma, Phys. Fluids **29**, 128 (1986).
[49] M. Mishra, R. Chhabra, and S. Sharma, Phys. Plasmas **1**, 70 (1994).

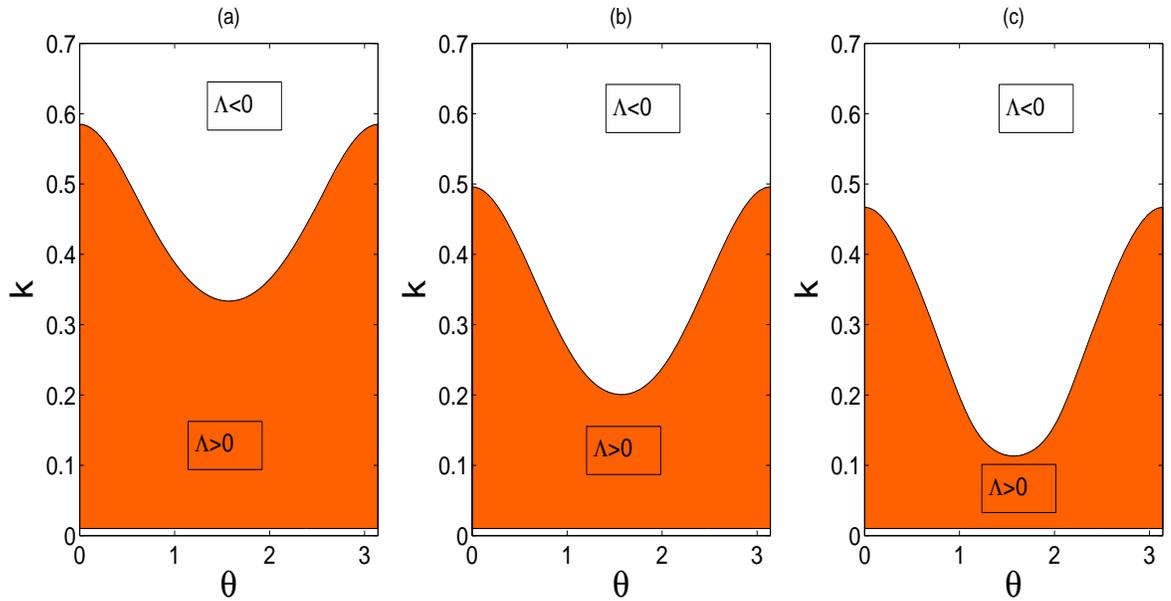

Fig. 1: Contour plots of $\Lambda = 0$ to show the stable $(\Lambda < 0)$ and unstable $(\Lambda > 0)$ regions in the $(k,\theta)$ plane for different values of $V_{th}$: (a) $V_{th} = 0.2c$, (b) $V_{th} = 0.1c$ and (c) $V_{th} = 0.05c$ with a fixed $k = 0.1$.

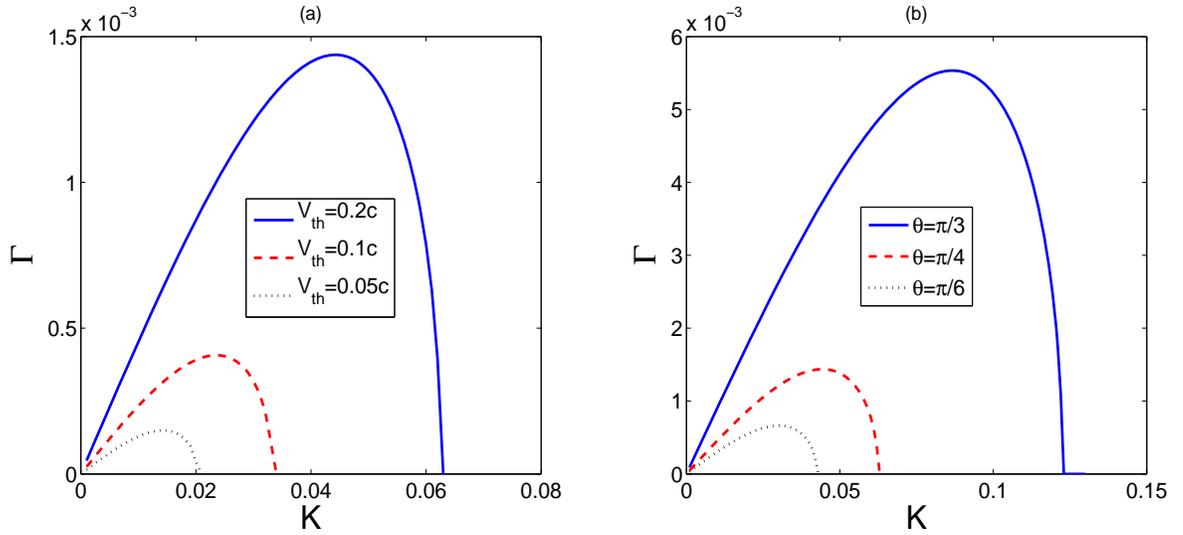

Fig. 2: The modulational instability growth rate $\Gamma$ is shown against the wave number of modulation $K$ for different values of (a) $V_{th}$ and (b) $\theta$ as in the legends. It is seen that the growth rates can be suppressed by reducing these parameter values.

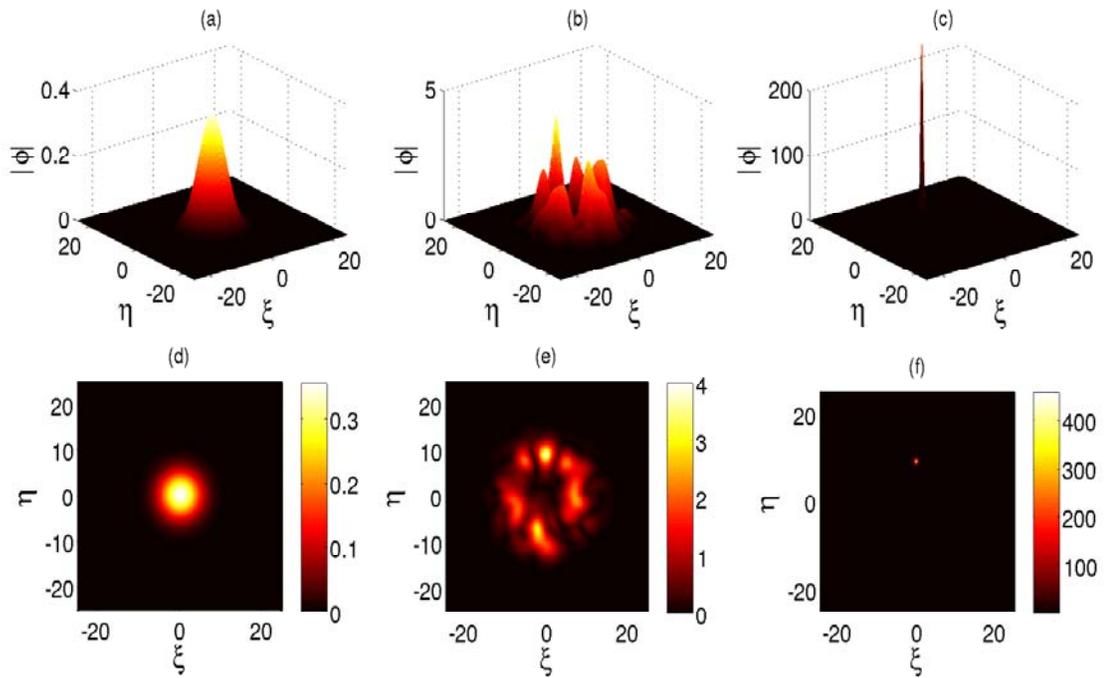

Fig. 3: Contour plots of a numerical solution of Eqs. (15) and (16) at different times (a) $\tau = 0$ (b) $\tau = 3.53$ and (c) $\tau = 3.56$ with a fixed $V_{th} = 0.2c$ and $k = 0.1$. The initial wave form (Gaussian) self-focuses and blows-up in a finite time.

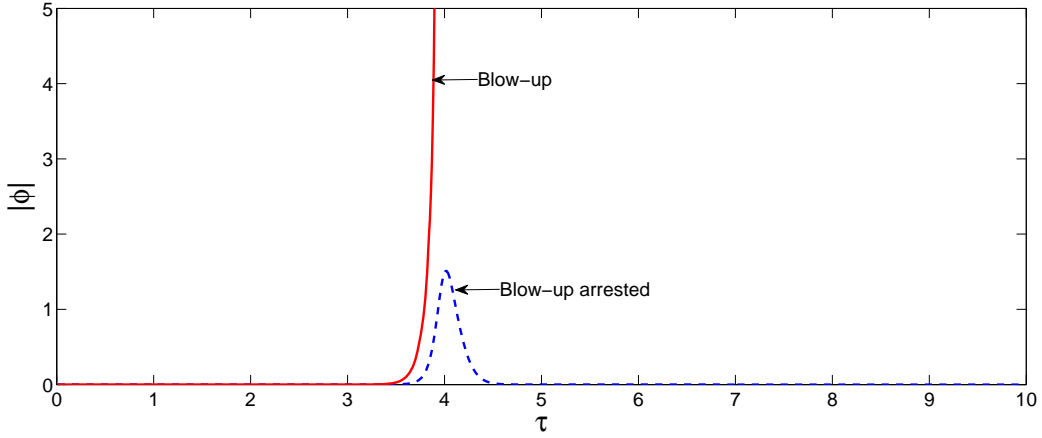

Fig. 4: Time evolution [numerical solution of Eqs. (15) and (16)] of the wave amplitude $\phi$ at different $V_{th}$: $V_{th} = 0.2c$ (solid line) and $V_{th} = 0.1c$ (dashed line) with a fixed $k = 0.1$. The solid line shows that the solution blows-up in a short time, while the dashed line indicates that the wave collapse is arrested (or singularity removed) with stable oscillations.

## Appendix A

$$P_1 = \frac{3v_{th}^2}{4\omega}\left[1 - \frac{3}{2}\frac{k_x^2 v_{th}^2}{\omega^2}\right],$$

$$P_2 = \frac{3v_{th}^2}{4\omega}\left[1 - \frac{3}{2}\frac{k_y^2 v_{th}^2}{\omega^2}\right],$$

$$P_3 = -\frac{3k_x k_y v_{th}^4}{4\omega^3},$$

$$Q_1 = \frac{3k^4 e^2 \omega}{4m_\circ^2 \omega_{pe}^4}\left(1 + \frac{v_{th}^2}{2c^2} + \frac{2\omega^2}{k^2 c^2}\right) + \frac{3Ak^2 v_{th}^2}{4\omega}\left(\frac{\omega^2}{k^2 c^2} - 1\right) - 2A\omega - \frac{2\omega}{3v_{th}^2}(\delta - \beta),$$

$$Q_2 = -2k_x,\ Q_3 = -2k_y,\ Q_4 = \frac{2\omega}{3v_{th}^2}\kappa,$$

$$R_1 = \frac{\beta - \delta}{v_{gx}}, \quad R_2 = \frac{\kappa}{v_{gx}}, \quad R_3 = \frac{k^2 e^2}{m_\circ^2 \omega_{pe}^2},$$

$$A = \frac{k^4 e^2 (4\omega^2 c^2 + 3\omega^2 v_{th}^2 + 3k^2 c^2 v_{th}^2)}{3 m_\circ^2 \omega_{pe}^2 c^2 (\omega_{pe}^2 + k^2 v_{th}^2)},$$

$$\alpha = \frac{3 k^4 v_{th}^2 e^2}{2 m_\circ^2 \omega_{pe}^4} + \frac{k_x^2 e^2}{m_\circ^2 \omega_{pe}^2} \left[ \frac{\omega^2}{\omega_{pe}^2} - \frac{3 k^2 v_{th}^4}{4 \omega_{pe}^2 c^2} \right],$$

$$\kappa = \frac{e^2 k_x k_y}{m_\circ^2 \omega_{pe}^2} \left[ \frac{\omega^2}{\omega_{pe}^2} - \frac{3 k^2 v_{th}^4}{4 \omega_{pe}^2 c^2} \right],$$

$$\beta = \frac{3 k^4 v_{th}^2 e^2}{2 m_\circ^2 \omega_{pe}^4} + \frac{k_y^2 e^2}{m_\circ^2 \omega_{pe}^2} \left[ \frac{\omega^2}{\omega_{pe}^2} - \frac{3 k^2 v_{th}^4}{4 \omega_{pe}^2 c^2} \right],$$

$$\delta = \frac{3 k^4 v_{th}^2 e^2}{2 m_\circ^2 \omega_{pe}^4} + \frac{k^2 e^2}{m_\circ^2 \omega_{pe}^2} \left[ \frac{\omega^2}{\omega_{pe}^2} - \frac{3 k^2 v_{th}^4}{4 \omega_{pe}^2 c^2} \right].$$

$$A_1 = n_0 A, \quad A_2 = -\frac{\pi e n_0}{k^2} A, \quad A_3 = \omega A - \frac{\omega}{2 n_0^2} \left( \frac{k^2}{4 \pi e} \right)^2.$$